\documentclass[pdftex,twocolumn,epjc3,a4paper]{svjour3}

\usepackage[switch]{lineno}

\RequirePackage[T1]{fontenc}
\smartqed  
\usepackage{amsmath}
\usepackage{amssymb}
\RequirePackage{flushend}
\RequirePackage{mathrsfs}
\RequirePackage{dsfont}
\usepackage{ulem}
\usepackage{graphicx}
\usepackage[colorlinks = true,
            linkcolor = blue,
            urlcolor  = blue,
            citecolor = blue,
            anchorcolor = blue]{hyperref}
\usepackage{cite}
\usepackage[utf8]{inputenc}
\usepackage{xcolor} 
\usepackage{upgreek}
\usepackage[switch]{lineno}
\journalname{Eur. Phys. J. C}
\usepackage{caption}
\usepackage{subcaption}
\usepackage{siunitx}

\newcommand{\iron}{$^{55}\text{Fe }$}

\begin{document}

\author{
  G.~Angloher\thanksref{addrMPI}\and
  S.~Banik\thanksref{addrHEPHY,addrAI}\and
  G.~Benato\thanksref{addrLNGS}\and
  A.~Bento\thanksref{addrMPI,addrCoimbra}\and 
  A.~Bertolini\thanksref{addrMPI,t1}\and 
  R.~Breier\thanksref{addrBratislava}\and
  C.~Bucci\thanksref{addrLNGS}\and 
  J.~Burkhart\thanksref{addrHEPHY}
  L.~Canonica\thanksref{addrMPI,addrINFNMilano,t2}\and 
  A.~D'Addabbo\thanksref{addrLNGS}\and
  S.~Di~Lorenzo\thanksref{addrMPI}\and
  L.~Einfalt\thanksref{addrHEPHY,addrAI}\and
  A.~Erb\thanksref{addrTUM,addrWMI}\and
  F.~v.~Feilitzsch\thanksref{addrTUM}\and 
  S.~Fichtinger\thanksref{addrHEPHY}\and
  D.~Fuchs\thanksref{addrMPI}\and 
  A.~Garai\thanksref{addrMPI}\and 
  V.M.~Ghete\thanksref{addrHEPHY}\and
  P.~Gorla\thanksref{addrLNGS}\and
  P.V.~Guillaumon\thanksref{addrLNGS}\and
  S.~Gupta\thanksref{addrHEPHY}\and 
  D.~Hauff\thanksref{addrMPI}\and 
  M.~Ješkovsk\'y\thanksref{addrBratislava}\and
  J.~Jochum\thanksref{addrTUE}\and
  M.~Kaznacheeva\thanksref{addrTUM}\and
  A.~Kinast\thanksref{addrTUM}\and
  H.~Kluck\thanksref{addrHEPHY}\and
  H.~Kraus\thanksref{addrOxford}\and 
  S.~Kuckuk\thanksref{addrTUE}\and
  A.~Langenk\"amper\thanksref{addrMPI}\and 
  M.~Mancuso\thanksref{addrMPI}\and
  L.~Marini\thanksref{addrLNGS,addrGSSI}\and 
  B.~Mauri\thanksref{addrMPI}\and
  L.~Meyer\thanksref{addrTUE}\and
  V.~Mokina\thanksref{addrHEPHY}\and
  M.~Olmi\thanksref{addrLNGS}\and
  T.~Ortmann\thanksref{addrTUM}\and
  C.~Pagliarone\thanksref{addrLNGS,addrCASS}\and
  L.~Pattavina\thanksref{addrLNGS,addrTUM}\and
  F.~Petricca\thanksref{addrMPI}\and 
  W.~Potzel\thanksref{addrTUM}\and 
  P.~Povinec\thanksref{addrBratislava}\and
  F.~Pr\"obst\thanksref{addrMPI}\and
  F.~Pucci\thanksref{addrMPI}\and 
  F.~Reindl\thanksref{addrHEPHY,addrAI} \and
  J.~Rothe\thanksref{addrTUM}\and 
  K.~Sch\"affner\thanksref{addrMPI}\and 
  J.~Schieck\thanksref{addrHEPHY,addrAI}\and 
  S.~Sch\"onert\thanksref{addrTUM}\and 
  C.~Schwertner\thanksref{addrHEPHY,addrAI}\and
  M.~Stahlberg\thanksref{addrMPI}\and 
  L.~Stodolsky\thanksref{addrMPI}\and 
  C.~Strandhagen\thanksref{addrTUE}\and
  R.~Strauss\thanksref{addrTUM}\and
  I.~Usherov\thanksref{addrTUE}\and
  F.~Wagner\thanksref{addrHEPHY}\and 
  M.~Willers\thanksref{addrTUM}\and 
  V.~Zema\thanksref{addrMPI}
(CRESST Collaboration)
}

\institute
{Max-Planck-Institut f\"ur Physik, D-80805 M\"unchen, Germany \label{addrMPI} \and
Institut f\"ur Hochenergiephysik der \"Osterreichischen Akademie der Wissenschaften, A-1050 Wien, Austria\label{addrHEPHY} \and
Atominstitut, Technische Universit\"at Wien, A-1020 Wien, Austria \label{addrAI} \and
INFN, Laboratori Nazionali del Gran Sasso, I-67100 Assergi, Italy \label{addrLNGS} \and
Comenius University, Faculty of Mathematics, Physics and Informatics, 84248 Bratislava, Slovakia \label{addrBratislava} \and
Physik-Department, TUM School of Natural Sciences, Technische Universität München, D-85747 Garching, Germany \label{addrTUM} \and
Eberhard-Karls-Universit\"at T\"ubingen, D-72076 T\"ubingen, Germany \label{addrTUE} \and
Department of Physics, University of Oxford, Oxford OX1 3RH, United Kingdom \label{addrOxford} \and
also at: LIBPhys-UC, Departamento de Fisica, Universidade de Coimbra, P3004 516 Coimbra, Portugal \label{addrCoimbra} \and
also at: Walther-Mei\ss ner-Institut f\"ur Tieftemperaturforschung, D-85748 Garching, Germany \label{addrWMI} \and
also at: GSSI-Gran Sasso Science Institute, I-67100 L'Aquila, Italy \label{addrGSSI} \and
also at: Dipartimento di Ingegneria Civile e Meccanica, Università degli Studi di Cassino e del Lazio Meridionale, I-03043 Cassino, Italy\label{addrCASS} \and
Present Address: INFN, Sezione di Milano LASA, Via Fratelli Cervi 201, I-20054 Segrate, Milano, Italy\label{addrINFNMilano} 
}

\thankstext[$\star$]{t1}{Corresponding author: anbertol@mpp.mpg.de}
\thankstext[$\dagger$]{t2}{Corresponding author: canonica@mpp.mpg.de}

\title{Light Dark Matter Search Using a Diamond Cryogenic Detector}

\date{Received: date / Accepted: date}

\maketitle

\begin{abstract}
Diamond operated as a cryogenic calorimeter is an excellent target for direct detection of low-mass dark matter candidates. Following the realization of the first low-threshold cryogenic detector that uses diamond as absorber for astroparticle physics applications, we now present the resulting exclusion limits on the elastic spin-independent interaction cross-section of dark matter with diamond. We measured two 0.175$\,$g CVD (Chemical Vapor Deposition) diamond samples, each instrumented with a W-TES. Thanks to the energy threshold of just 16.8$\,$eV of one of the two detectors, we set exclusion limits on the elastic spin-independent interaction of dark matter particles with carbon nuclei down to dark matter masses as low as 0.122$\,$GeV/c$^2$. 
This work shows the scientific potential of cryogenic detectors made from diamond and lays the foundation for the use of this material as target for direct detection dark matter experiments. 


\end{abstract}

\section{Introduction}
Dark matter (DM) is one of the most investigated topics in astroparticle physics. Its presence is highly motivated by many observational evidences \cite{Salucci2018, Massey_2010, planck}. Many theories have been built around the idea of a particle-like DM, predicting candidates that cover an extended mass range. In the last decade cryogenic experiments have been very successful in reaching extremely low energy thresholds, taking on a crucial role in the exploration of DM in the GeV mass range and below. Among them, the Cryogenic Rare Event Search with Superconducting Thermometers (CRESST) experiment has recently gained sensitivity to DM masses of 0.115$\,$GeV/c$^2$ with an energy threshold down to 10$\,$eV in an underground measurement \cite{silicon}.\\
We have reported in a previous publication how a similar energy threshold was achieved in an above ground measurement employing diamond single crystals as detector material\cite{diamond1}. In this work we report limits on the elastic spin-independent DM-nucleon interactions using data obtained with these detectors. The experimental setup, data taking and energy calibration will be described very concisely in this work. For a more detailed description, we refer the reader to our previous work in \cite{diamond1}.

\begin{figure*}[!t]
    \centering
    \includegraphics[width=\textwidth]{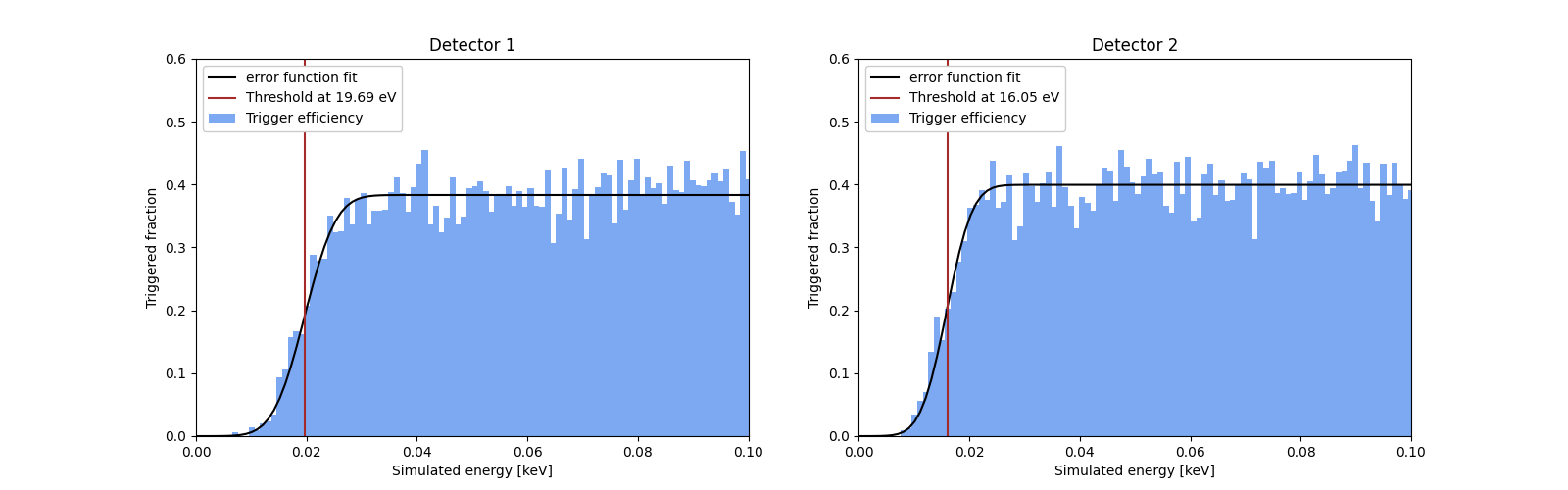}
    \caption{Confirmation of the energy threshold of detector 1 and 2 using simulated data. We plotted the trigger efficiency against the simulated energy value (blue) and fitted the distribution with an error function (black line). We expect the threshold value to be at the energy value where the trigger efficiency drops below half of the constant trigger efficiency at higher energies. For detector 1 the threshold value is (19.7 $\pm$ 5.1)$\,$eV while for detector 2 it is (16.1 $\pm$ 4.4)$\,$eV (brown lines). For both detectors the threshold of the previous publication of 19.7$\,$eV and 16.8$\,$eV are within the fit errors.}
    \label{fig:threshold_Ernie}
\end{figure*}

\section{Diamond as cryogenic detectors}
Cryogenic calorimeters are used in many different fields of astroparticle physics. See \cite{PirroNTD} for a comprehensive review. The high interest in these devices can be attributed among others to the possibility of using different materials as energy absorbers, with the remarkable advantage that the most suitable material can be chosen depending on the particular research purpose\cite{PRETZL2000114}. \\
With a Debye temperature of 2220$\,$K and therefore a favorable phonon propagation, diamond crystals have the properties to be excellent absorbers for cryogenic calorimeters aiming at reaching low-energy thresholds. Additionally, the light nucleus of carbon (A=12) allows to probe lower DM masses, being kinematically favored compared to heavier target nuclei. A more detailed description of the advantages of using diamond as cryogenic DM detectors can be found in \cite{Kurinsky}.\\
In \cite{diamond1} we describe the experimental setup realized to operate two diamond single crystals of 0.175$\,$g and a size of (2x5x5)$\,$mm$^3$ each. In this work they will be referred to as \textit{detector 1} and \textit{detector 2}. Each of them has been instrumented with a W-TES and operated in a dilution refrigerator at the Max-Planck-Institute for Physics in Munich, Germany, in an above ground facility without radiation shielding. Both detectors achieved an excellent performance, reaching a baseline resolution of 3.54$\,$eV and 3.42$\,$eV respectively and energy thresholds of 19.7$\,$eV and 16.8$\,$eV, derived with the method described in \cite{diamond1}.

\begin{figure*}[!t]
    \centering
    \includegraphics[width=\textwidth]{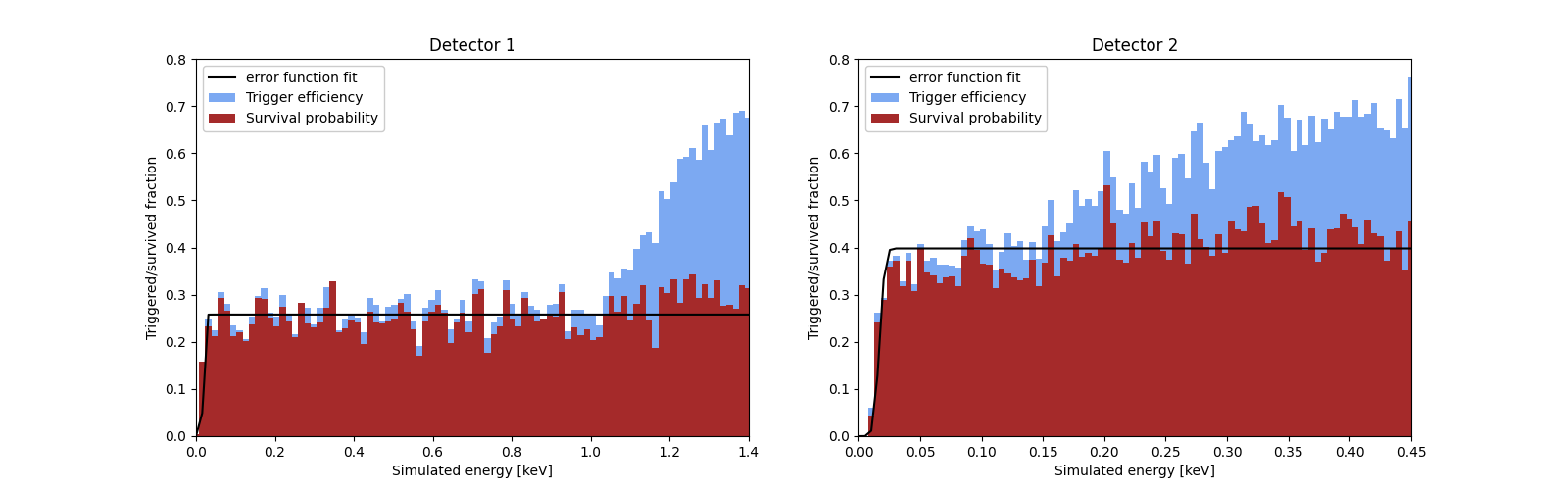}
    \caption{Trigger efficiency (blue) and survival probability (brown) of detector 1 and 2. We calculated the probability of simulated events surviving the trigger and the quality cuts. The distribution of the survival probability has been fitted with an error function (black line). The constant survival probability of detector 1 is 25.8 \% while of detector 2 it is 39.8 \%.}
    \label{fig:surv_prob}
\end{figure*}

\section{Data Analysis}

\subsection{Data processing}
\sloppy
The data has been processed and calibrated using the same procedure described in our previous publication\cite{diamond1}. In order to optimize the trigger threshold we adopted an offline triggering method.  For this approach we recorded the complete stream of data and then processed it offline with an optimum filter \cite{OptimumFilter_GM}. This was created from the noise power spectrum of the specific noise conditions of the measurements, and from the shape of an averaged particle event, also called standard event. \\
During data processing, the data stream is first divided into windows of 655 ms around the triggered timestamp (in case of multiple events in the same time window, the highest pulse in the window is set at the correct position) and then some parameters that describe the shape of the pulses and of the baseline (e.g. pulse height, difference of the average baseline values at the beginning and at the end of the window and RMS of the baseline) are calculated.\\

\subsection{Event Selection}
In the analysis procedure, we used the same filter used for the offline triggering also for the amplitude estimation at the trigger position. We extracted additional parameters from the filtered data, in particular, the amplitude value evaluated by the filter and the filter RMS which quantifies the difference between the filtered pulse and the filtered standard event. The latter determines the deviation of the particle pulse shape from the one of the standard event that was used to create the optimum filter. Using all these parameters, we applied several cuts to remove artifacts and to select only windows where we could assure a correct amplitude reconstruction of the pulse.\\
The main artifacts in our data set were caused by the fast rise time of the pulses. The readout electronic was too slow for high energetic pulses and caused a reset of the baseline with respect to the pre-trigger range which deformed the pulse shape. Given that these artifacts have different baseline values before and after the reset, they were easily removed by selecting only events with a small difference of the average of the baseline at the beginning and at the end of the window.\\
We also accepted only those events with the best noise condition and therefore discarded events with a high baseline RMS. Finally, we excluded remaining artifacts and distorted pulse shapes by applying a cut on the ratio of the filter RMS and the filter amplitude.\\
We removed several hours at the beginning of the data taking where the detector response was very unstable. The stability check is performed by injecting heater pulses throughout the whole measurement with the purpose of monitoring the detector response over time (see \cite{diamond1}). After this stability cut, our final data set counted 37.08 h measuring time that resulted in an exposure of 0.27$\,$g$\cdot$d.\\
Finally, the acquired data were calibrated using X-rays of 5.89$\,$keV and 6.49$\,$keV emitted by an \iron source that was located inside the detector holder.

\subsection{Trigger efficiency and survival probability}
\begin{figure}
    \centering
    \includegraphics[width=.49\textwidth]{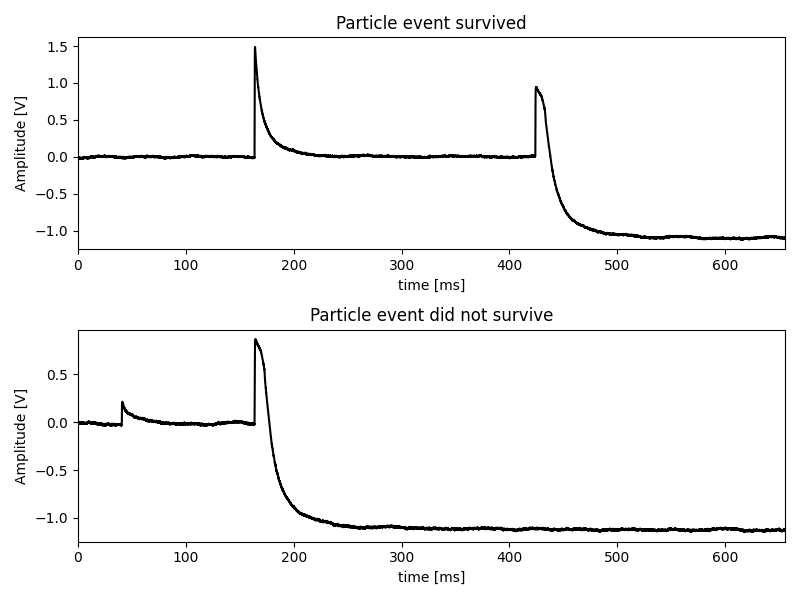}
    \caption{Visualization of the baseline reset after a high energetic event. In both plots a simulated signal event is followed by a baseline reset caused by an energetic particle. In the upper plot the simulated event is reconstructed at the correct onset (164 ms) and tagged as triggered because its amplitude is higher than the artifact, while in the lower plot the artifact is higher and therefore the simulated event is not tagged as triggered.
    \label{fig:artifact}}
\end{figure}

\begin{figure}
    \centering
    \includegraphics[width=.49\textwidth]{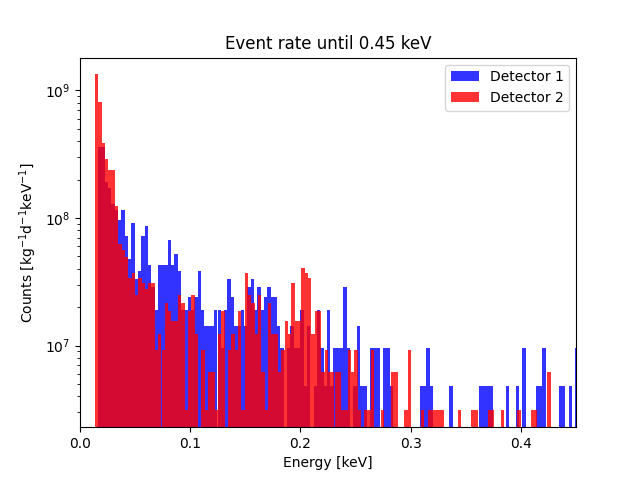}
    \caption{Event rate per kg day keV corrected with survival probability. We plotted the calibrated spectrum after trigger and quality cuts and corrected each energy bin of the size of 3 eV with the constant survival probability. Detector 2 (red) is plotted until the end of its dynamic range (0.45$\,$keV) while detector 1 (blue) is plotted only until 0.45$\,$keV for a better visualization.
    \label{fig:spectrum}}
\end{figure}

Once we obtained a calibrated energy spectrum for each detector, we also performed a simulation to estimate the signal survival probability, i.e. with which probability valid signal events survive the data processing and cleaning steps. For this purpose we simulated particle-like events with a flat energy spectrum from 0 until the end of the dynamic range of each detector which is 1.4$\,$keV for detector 1 and 0.45$\,$keV for detector 2.\\
These events were simulated by superimposing scaled standard particle events on our real data stream at random times. The voltage amplitude of each simulated pulse was determined with a specific time-dependent detector response function to account for the effect of instabilities. The detector response of each point in time could be studied with the heater pulses. By applying the identical analysis steps as for real data we studied the probability of signal events surviving the trigger algorithm and our quality cuts. To avoid an overestimation of the signal survival probability we removed events where the simulated and the reconstructed amplitude differed by more than 3 times the baseline resolution of the detector (with this cut we removed simulated events under threshold that coincided with strong upward fluctuations of the baseline). The result of the trigger and survival probability can be seen in figure \ref{fig:threshold_Ernie} and \ref{fig:surv_prob}.\\
In figure \ref{fig:threshold_Ernie}, we plotted only the trigger efficiency that we determined with a dedicated simulation in a limited energy range until 0.1$\,$keV to enhance the statistics at low energies. With this simulated data set we calculated the ratio of the triggered events to the total number of simulated events and fitted it with an error function. We expect the energy threshold to be at the simulated energy value where the error function drops below half of its constant value. With the fit we obtained the value of (19.7 $\pm$ 5.1)$\,$eV for detector 1 and (16.1 $\pm$ 4.4)$\,$eV for detector 2.
This confirms our previous energy threshold cited in \cite{diamond1} which was calculated by simply converting the voltage threshold into eV using a calibration factor.\\
For the determination of the trigger efficiency and the survival probability over the whole dynamic range of the two detectors we used a second set of simulated data. The results are presented in figure \ref{fig:surv_prob}. As can be seen, both detectors suffer from a very low trigger efficiency. This is due to the high event rate above ground which is accounted for at trigger level where in case of multiple pulses in the same time window only the largest one was tagged as triggered.\\
The trigger efficiency shows an energy dependence, not equally pronounced in the two detectors, due to an artifact caused by high energy particles. The fast rise of these pulses caused resets of the baseline that resulted in pulses being assigned a fixed wrong amplitude (calibrated at about 1.1$\,$keV in detector 1 and around 0.2$\,$keV in detector 2). Events smaller than this amplitude were hidden by this artifact and were therefore tagged as not triggered (see figure \ref{fig:artifact}). Such energy dependence is not present in the signal survival probability since pulses with the incorrect pulse shape are effectively removed.\\   
We fitted the survival probability with an error function considering a flat survival probability at high energies. With this fit we estimated a survival probability of 25.8\% in detector 1 and one of 39.8\% in detector 2.\\
Figure \ref{fig:spectrum} shows the final calibrated spectra for both detectors corrected with the corresponding survival probability. For better visualization both detectors are plotted up to the same energy value of 0.45$\,$keV, which corresponds to the end of the dynamic range of detector 2.

\vspace{5cm}

\section{Dark Matter Results}
\begin{figure}
    \centering
    \includegraphics[width=.49\textwidth]{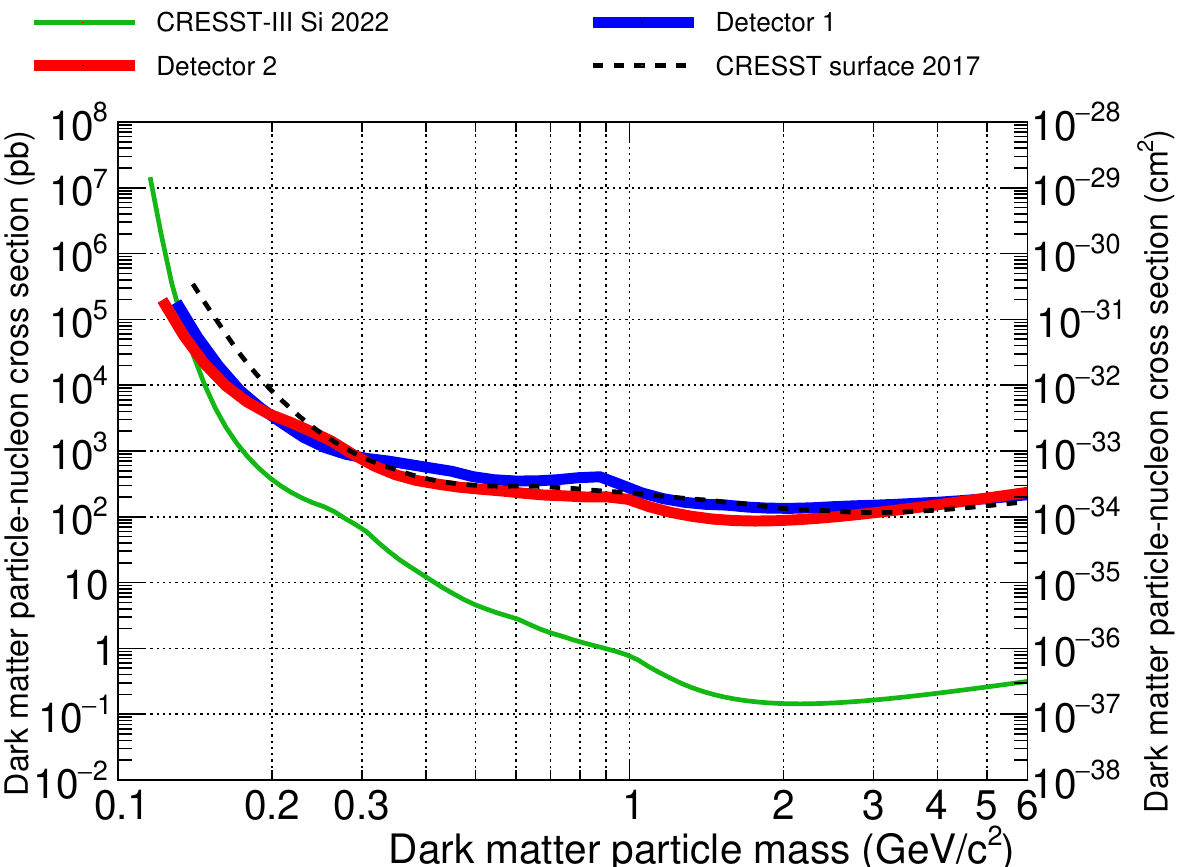}
    \caption{Exclusion limits for the elastic spin-independent DM-nucleon scattering cross section at 90\% CL, calculated for detector 1 (blue) and 2 (red) using Yellin's optimum interval method. In black, the previous best above ground exclusion limits of CRESST are plotted \cite{CRESSTabove17}. In green, the best exclusion limits below 0.160$\,$GeV/$c^2$ from CRESST underground measurements \cite{silicon} are plotted as a benchmark reference.  
     }
    \label{fig:limit}
\end{figure}

\begin{figure}
    \centering
    \includegraphics[width=.49\textwidth]{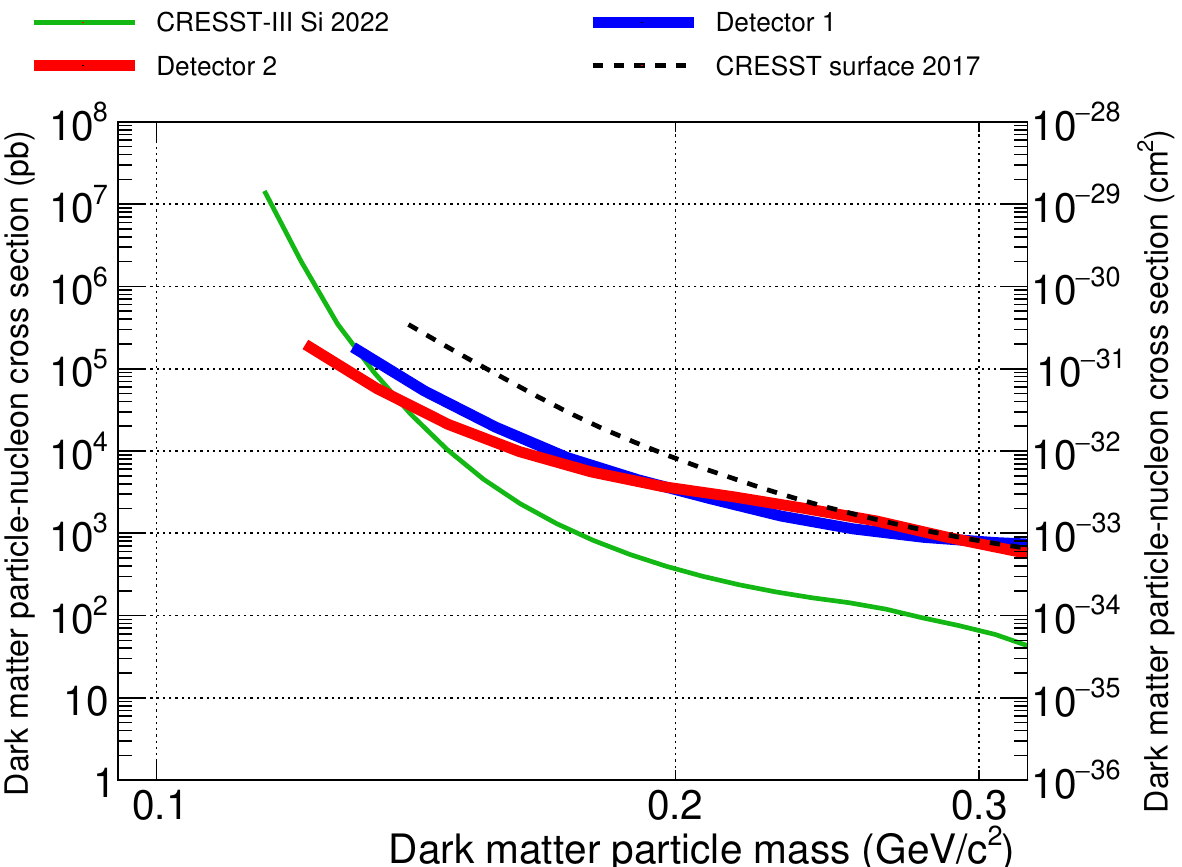}
    \caption{Zoomed in version of figure \ref{fig:limit}. From this picture it is more evident how detector 1 (blue) and 2 (red) are excluding additional parameter space compared to the previous best above ground limits. }
    \label{fig:limitzoom}
\end{figure}

\sloppy
The final energy spectra of our analysis (figure \ref{fig:spectrum}) show a rise of events towards low energies. This is an effect that is not new to the scientific community, as it has been observed in many other experiments operated both  underground and above ground, as described in \cite{excess}. We conservatively consider these events as potential signal and calculate exclusion limits adopting Yellin's optimum interval method \cite{Yellin_2002,Yellin_2007}. The upper limit on the elastic spin-independent DM-nucleon interaction is derived by comparing for each DM particle mass the observed spectrum with the expected one, corrected with the detector response as obtained by simulation. For the calculation of the expected differential energy spectra we adopted the standard DM halo model, with an asymptotic velocity of $\upsilon_\odot$ = 220~km/s \cite{solarvelocity}, a local DM density of $\rho_{\textup{DM}}$ = 0.3~(GeV/c\textsuperscript{2})/cm\textsuperscript{3} \cite{density} and the galactic escape velocity of $\upsilon_{\textup{esc}}$ = 544~km/s \cite{velocity}.\\
The resulting elastic spin-independent DM-nucleon scattering cross section exclusion limits with 90\% confidence level are shown in figure \ref{fig:limit}. A zoomed in version of it can be seen in figure \ref{fig:limitzoom}. In these plots we are comparing the exclusion limits obtained with the diamond detector only to the previous CRESST results, in order to highlight the potential of the use of this new material compared to standard CRESST detectors.\\
One can clearly observe that, due to the light nucleus, the diamond detectors are extending the excluded parameter space to lower DM masses compared to the previous best above ground limits of CRESST \cite{CRESSTabove17} (dashed black in figure \ref{fig:limit} and \ref{fig:limitzoom}), that was obtained using a 0.5$\,$g sapphire detector with an energy threshold of 19.7$\,$eV. Using the detector 2 results it was possible to exclude masses until 0.122$\,$GeV/c$^2$. For large dark matter masses the sensitivity of all the above ground measurements is limited by the low exposure and by the background. \\
The green curve shows as a reference the current best limit from CRESST for masses below 0.16$\,$GeV, which was obtained with a 0.35$\,$g silicon wafer detector with a threshold of 10$\,$eV in the well shielded underground setup of CRESST at the LNGS \cite{silicon}. The lower background in the below ground measurement leads to a much better limit at higher masses. At low masses the diamonds cover a similar range compared to the silicon results despite the higher threshold. This highlights again the advantage of using a material with light target nuclei and demonstrate the potential of using diamond as a target in cryogenic detectors for low mass direct dark matter searches.\\

\section{Conclusions}
These results demonstrate the potential of cryogenic detectors using diamond as target material for direct DM searches. In particular, their properties make them an ideal material for low-threshold experiments. With this proof-of-principle measurement we reach an energy threshold of 16.8$\,$eV on the best performing detector, which allows for a sensitivity to DM masses down to 0.122$\,$GeV/c$^2$.\\
Figures \ref{fig:limit} and \ref{fig:limitzoom} show how thanks to the lighter nucleus diamond could exclude a larger parameter space compared to the previous best above ground measurement which had a comparable threshold and exposure. The difference with respect to the best underground limit has to be attributed not only to the different mass of the nucleus but also on the differences in energy threshold, exposure and low energy background.\\ 
Diamond has the potential to be sensitive to a larger parameter space than the one presented in this work by pushing down the energy threshold and reduce the background in an underground measurement. Therefore, we are planning to extend our research with this material. In particular we aim to reach a higher exposure using larger crystals, and a better performance thanks to an improved read-out chain and an optimized W-TES sensor design. With these improvements, cryogenic diamond detectors will have the possibility to explore new properties for the interaction of sub-GeV DM with ordinary matter.

\begin{acknowledgements}
This research was supported by the Excellence Cluster ORIGINS which is funded by the Deutsche Forschungsgemeinschaft (DFG, German Research Foundation) under Germany's Excellence Strategy – EXC-2094 – 390783311.

\end{acknowledgements}

\bibliographystyle{modified_spphys}
\bibliography{biblio}

\end{document}